\documentclass[useAMS,usenatbib,usegraphicx]{mn2e}

\usepackage{times,epsfig}

\def\gsim{\mathrel{\raise0.35ex\hbox{$\scriptstyle >$}\kern-0.6em
\lower0.40ex\hbox{{$\scriptstyle \sim$}}}}
\def\lsim{\mathrel{\raise0.35ex\hbox{$\scriptstyle <$}\kern-0.6em
\lower0.40ex\hbox{{$\scriptstyle \sim$}}}}
\def\m@th{\mathsurround=0pt }
\def\eqalign#1{\null\,\vcenter{\openup1\jot \m@th
 \ialign{\strut\hfil$\displaystyle{##}$&$\displaystyle{{}##}$\hfil
 \crcr#1\crcr}}\,}
\def\micron{$\mu\textrm{m}$}
\def\micronend{$\mu\textrm{m}$}
\def\microjy{$\mu\textrm{Jy}$ }

\def\microjyend{$\mu\textrm{Jy}$}

\title[The nature of the brightest SMGs]
      {Clarifying the nature of the brightest submillimetre sources: interferometric imaging of LH850.02}
\author[Younger et al.]{
J.\,D.\ Younger,$^{\! 1}$\thanks{E-mail: jyounger@cfa.harvard.edu} 
J.\,S.\ Dunlop,$^{\! 2,3}$
A.\,B.\ Peck,$^{\! 1,4}$
R.\,J.\ Ivison,$^{\! 2,5}$
A.\,D.\ Biggs,$^{\! 5}$
E.\, L.\ Chapin,$^{\! 3}$ \and
D.\, L.\ Clements,$^{\! 6}$ 
S.\ Dye,$^{\! 7}$
T.\,R.\ Greve,$^{\! 8}$
D.\,H.\ Hughes,$^{\! 9}$
D.\ Iono,$^{\! 10}$
I. Smail, $^{\! 11}$
M.\ Krips,$^{\! 1}$ \and
G.\,R.\ Petitpas,$^{\! 1}$
D.\, Wilner,$^{\! 1}$
A.\,M.\ Schael,$^{\! 2}$
and C.\, D.\ Wilson$^{\! 12}$
\vspace*{1mm}\\
$^1$ Harvard-Smithsonian Center for Astrophysics, 60 Garden Street,
     Cambridge, MA 02138, USA\\
$^2$ SUPA (Scottish Universities Physics Alliance), 
Institute for Astronomy,
     University of Edinburgh, Royal Observatory, Blackford Hill, Edinburgh EH9 3HJ\\
$^3$ Department of Physics and Astronomy, University of British Columbia, 6224 Agricultural Road, Vancouver V6T 1Z1, Canada \\
$^4$ Joint ALMA Office, El Golf 40, Las Condes, Santiago 7550108, Chile\\
$^5$ UK Astronomy Technology Centre, Royal Observatory, Blackford Hill,
     Edinburgh EH9 3HJ\\
$^6$ Astrophysics Group, Blackett Laboratory, Imperial College, Prince Consort Road, 
     London SW7 2BW, UK \\
$^7$ Cardiff University, School of Physics \& Astronomy, Queens Buildings, The Parade, 
     Cardiff CF24 3AA\\
$^8$ Astronomy Department, Max--Planck--Institut f\"{u}r Astronomie, K\"{o}ningstuhl--17, D--69117,
     Heidelberg, Germany \\
$^9$ Instituto Nacional de Astrof\'{i}sica, \'{O}ptica y Electr\'{o}nica (INAOE), Tonantzintla, Peubla, M\'{e}xico \\
$^{10}$ National Astronomical Observatory of Japan (NAOJ), 2--21--1 Osawa, Mitaka,
     Tokyo 181--8588, Japan\\
$^{11}$ Institute for Computational Cosmology, Durham University, South Road,
     Durham DH1 3LE \\
$^{12}$ Department of Physics \& Astronomy, McMaster University, Hamilton, 
            Ontario, L8S 4M1, Canada \\
}

\date{\fbox{\sc Draft dated: \today\ }}

\pagerange{000--000}

\begin{document}

\maketitle

\begin{abstract}

We present high-resolution interferometric imaging of LH\,850.02, the
brightest 850- and 1200-\micron\ submillimetre (submm) galaxy in the
Lockman Hole. Our observations were made at 890\,\micron\ with the
Submillimetre Array (SMA). Our high-resolution submm imaging detects
LH\,850.02 at $\gsim$6$\sigma$ as a single compact (size
$\lsim$1\,arcsec or $\lsim$8\,kpc) point source and yields its
absolute position to $\sim$0.2-arcsec accuracy. LH\,850.02 has two alternative 
radio counterparts within the SCUBA beam (LH850.02N \& S), both of which 
are statistically very unlikely to be so close to the SCUBA source 
position by chance. However, the precise astrometry from the SMA 
shows that the submm 
emission arises entirely from LH850.02N, and is {\it not} associated 
with LH850.02S (by far the brighter of the two alternative identifications
at 24--\micron\ ). Fits to the optical-infrared multi-colour photometry of 
LH850.02N \& S indicate that both lie at $z\approx 3.3$, and are therefore
likely to be physically associated.  
At these redshifts, the 24\micronend--to--submm flux density ratios suggest 
that LH\,850.02N has an Arp\,220-type starburst-dominated far-IR SED, 
while LH\,850.02S is more similar to Mrk\,231, with less dust-enshrouded 
star-formation activity, but a significant contribution at 
24--\micron\ (rest-frame ~$5-6$\, \micron ) from an active nucleus.  
This complex mix of star--formation and AGN activity in multi--component sources may be 
common in the high redshift ultraluminous galaxy population, and highlights the need for  precise astrometry  from high resolution interferometric imaging for a more complete understanding.

\end{abstract}


\section{Introduction}

\begin{figure*}
\epsfig{figure=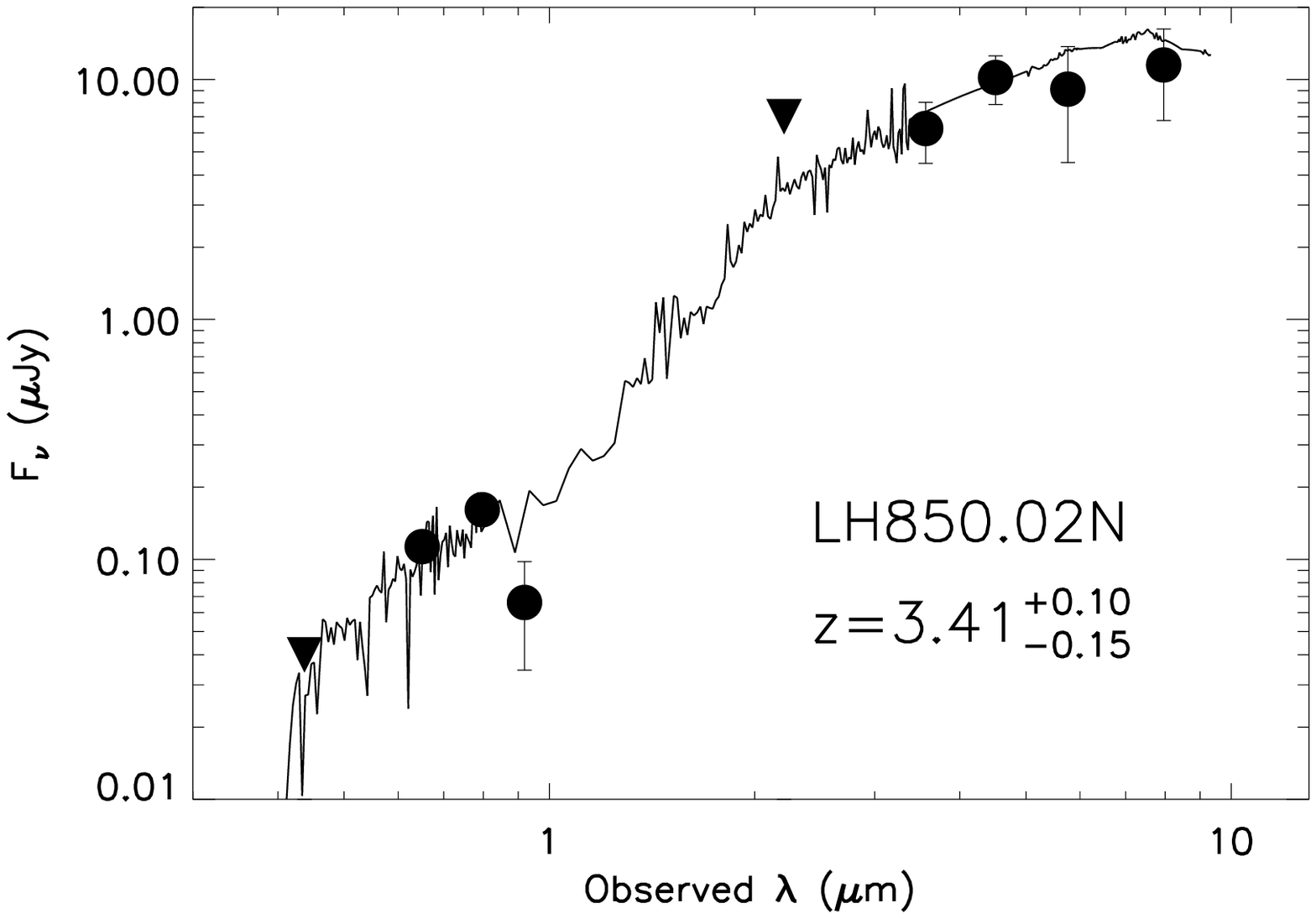,width=80mm}
\epsfig{figure=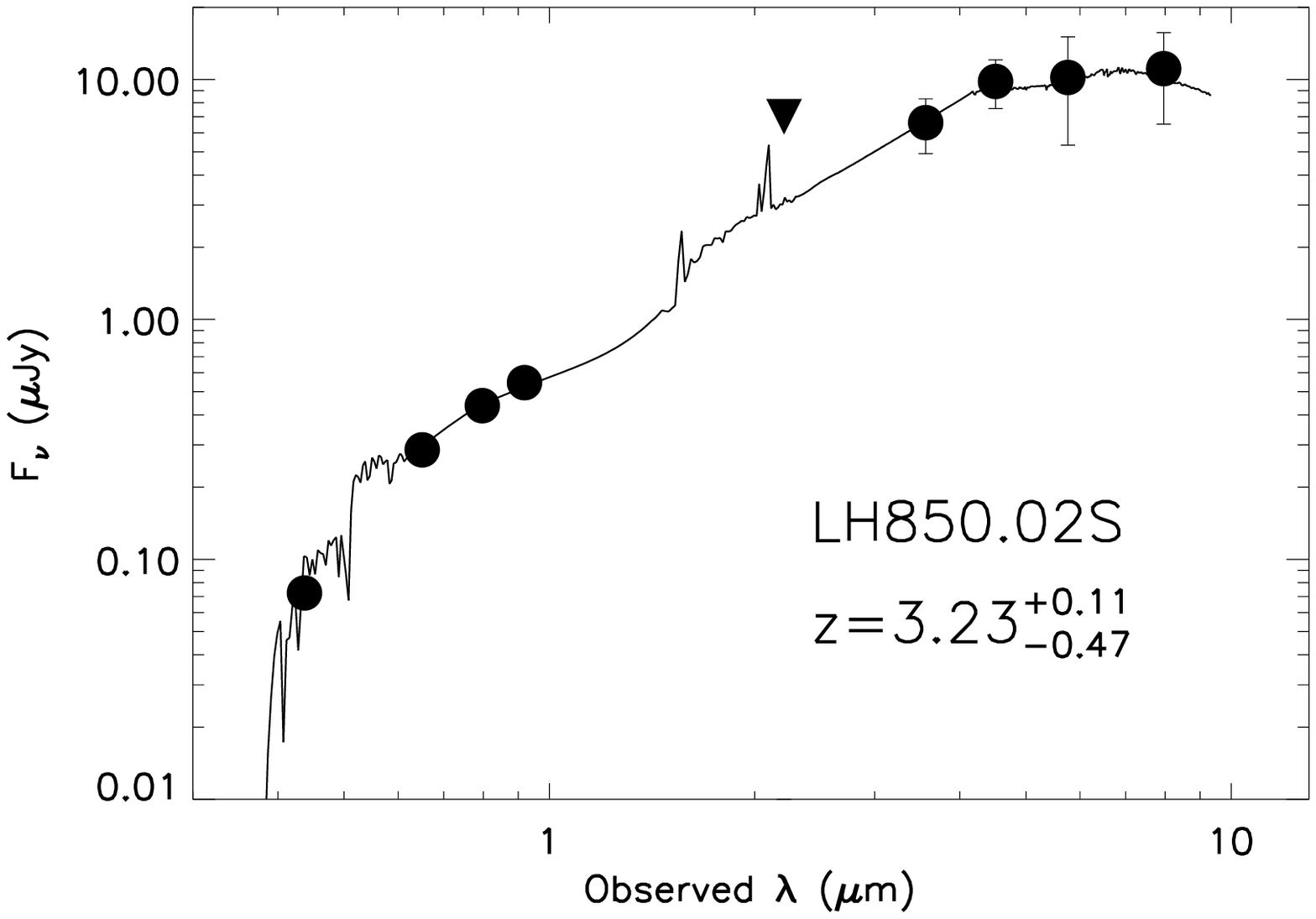,width=80mm}
\caption{Optical and near-IR photometry for LH\,850.02N and
LH\,850.02S, along with fits to their spectral energy distributions
(SEDs), and photometric redshifts \citep{dye2008}.  Downward arrows
indicate upper limits.}
\label{fig:photoz}
\end{figure*} 

It has been well established that up to half of the far-infrared
(far-IR) extragalactic background is produced by dusty starbursts and
active galactic nuclei \citep{hauser1998,dwek1998,fixsen1998,pei1999}.
A significant fraction of this background was resolved
at 850\,\micron\ into discrete point sources
\citep{smail1997,hughes1998,barger1998} by the Submm Common-User
Bolometric Array \citep[SCUBA:][]{holland1999} on the 15-m James Clerk
Maxwell Telescope (JCMT).  These so-called submm galaxies (SMGs) are
thought to be high-redshift ultraluminous and hyperluminous IR
galaxies \citep[see][]{chapman2005} that represent massive systems in
the process of formation \citep{scott2002,blain2004}, and may dominate
cosmic star formation for nearly the first half of the lifetime of the
Universe \citep[$z\gsim 1$;][]{blain1999,blain2002}.

Since their discovery a decade ago, a series of 
surveys using SCUBA at 850\,\micron\ \citep{barger1999,eales1999,eales2000,
chapman2002,cowie2002,scott2002,borys2002,webb2003,borys2003,serjeant2003,wang2004,knudsen2006,coppin2006,knudsen2008} and mm
instruments at longer wavelengths \citep{greve2004,dannerbauer2004,
laurent2005,dscott2006,bertoldi2007,scott2008} have
amassed catalogues of hundreds of SMGs. Unfortunately, the detailed
study of these objects has been hindered somewhat by the poor angular
resolution ($\sim$11--18\,arcsec) of current submm/mm telescopes.  
This problem was first
addressed via deep radio continuum surveys, which exploited the
radio--FIR correlation \citep[see][for a review]{condon1992} in
combination with statistical arguments \citep[e.g.,][]{ivison2002,
ivison2007} to associate 1.4-GHz sources with
the submm emission.  Radio counterparts allowed SMGs to be localised
with sub-arcsec precision and hence allowed a more detailed study of their
properties. Optical spectroscopy of these radio-identified
samples confirmed that SMGs lie preferentially at high redshift
\citep[median $z\sim 2.3$;][]{chapman2005} and enabled CO spectroscopy
which revealed them to be compact, massive, gas-rich, possibly merging
systems \citep[e.g.,][]{neri2003,sheth2004,kneib2005,greve2005,tacconi2006}.

Despite the undoubted success of deep radio imaging of SMGs, there is still
a clear need for high-resolution {\it submillimetre/millimetre} observations of at least 
a subset of the SMGs found in current, complete 
samples. Specifically, two classes of object require such observations 
to locate the source of the submillimetre emission with the required 
astronomical precision for optical/infrared follow-up. 

First, because of the rapid dimming of the radio continuum with
redshift \citep[$I_\nu\sim (1+z)^{\alpha-3}$, $\alpha=-0.8$, where
$I_{\nu}\propto\nu^{\alpha}$;][]{condon1992}, even the deepest
existing radio imaging is relatively insensitive to SMGs at $z\gsim
3$; as a result, typically only around two thirds of SMGs have been
detected at radio wavelengths \citep{ivison2002}.  Other techniques
\citep{ashby2006,pope2006} have been suggested, which make use of data
from the IR Array Camera \citep[IRAC:][]{fazio2004}, in combination
with 24-\micron\ observations using the Multiband Imaging Photometer
\citep[MIPS:][]{rieke2004}, on board the {\it Spitzer Space Telescope}
to select counterparts.  However, they too may have hidden biases
which are difficult to quantify, and the
only unambiguous way to select the correct optical 
counterparts for these radio-unidentified SMGs is via
time-intensive submm/mm high-resolution interferometric imaging. 
Interferometric observations at
mm \citep{downes1999,frayer2000,dannerbauer2002,genzel2003,
greve2005,tacconi2006,dannerbauer2008} and submm \citep[][see also Iono et al. 2008, in prep.]{iono2006,wang2007}
wavelengths have successfully detected a growing catalogue of SMGs
(in the process also confirming the reliability of the radio--submm 
association where a unambiguous radio counterpart has already been discovered).
Most recently, \citet{younger2007} followed up a flux-limited sample of seven
mm-selected SMGs detected by the AzTEC camera \citep{wilson2008} at
the JCMT -- including five without reliable radio identifications --
at 890\,\micron\ with the Submillimeter Array \citep[SMA:][]{ho2004},
the counterparts of which suggested a population of very luminous SMGs
at higher redshift than radio-identified samples (see also Yun et al. 2008, in prep.).
That this bright (median 890-\micron\ flux density 12.0\,mJy), AzTEC-selected 
sample contains a significant high-redshift tail 
of SMGs supports earlier evidence that the brightest SMGs may be
the most distant \citep[e.g.\ Figure~9 of][]{ivison2002,dunlop2001}.

Second, within the radio-identified sub-samples of SMGs there remains 
some confusion. Specifically, there exists a 
statistically significant fraction of SMGs which possess more than one 
statistically robust radio counterpart \citep[$\sim20\pm5\%$;][]{ivison2007}.  
Monte--carlo simulations suggest that these associations are observed 
significantly more frequently than would be expected from chance associations.  
The nature of these multiply-identified SMGs remains somewhat uncertain, although 
the steepness of the SMG number counts suggests that the bright submillimetre 
sources are only rarely expected to arise from the blending/confusion 
of more moderate luminosity subcomponents. Interferometric imaging of the 
rest--frame far--IR continuum offers the best way to investigate 
the true nature of these interesting systems by 
precisely locating the source of the submm emission 
\citep[see e.g., SMM J094303+4700 H6 and H7;][]{tacconi2006}.

In this paper, we present high-resolution SMA 890-\micron\
interferometric imaging of LH\,850.02/LH\,1200.04 (hereafter simply
LH\,850.02), the brightest source in both the MAMBO 1200-\micron\
\citep{greve2004} and SCUBA 850-\micron\ \citep{coppin2006} maps of
the Lockman Hole.  In \S~\ref{sec:obs} we describe our
observations and data reduction, and in \S~\ref{sec:discuss} we
discuss some possible implications of our results.  Throughout this
paper, we use the Vega magnitude system, and assume a flat concordance
cosmology with $\rm (\Omega_m,\Omega_\Lambda,H_0) = (0.3,0.7,70$
km\,s$^{-1}$\,Mpc$^{-1}$).

\section{Target Selection}
\label{sec:target}

SHADES is a wide-area blank-field submm survey, covering both the Lockman Hole (LH) 
and Subaru/{\em XMM--Newton} Deep Field (SXDF), undertaken using SCUBA on 
the JCMT between 2002 and 2005 \citep{dunlop2005,mortier2005}.  The map of the LH region covers an area of 
485\,arcmin$^2$, mostly to an r.m.s.\ noise level of 2.2\,mJy\,beam$^{-1}$. 
The field has extensive complementary observations, including 1.4-GHz radio 
continuum imaging from the Very Large Array (VLA) to an r.m.s.\ noise level of 
4.2\,\microjy\,beam$^{-1}$ \citep{ivison2002,biggs2006,ivison2007}, 
{\em Spitzer}/IRAC and MIPS 3.6- and 24-\micron\ imaging to $5\sigma$ 
sensitivity limits of 1.3 and 55\,\microjy, respectively 
\citep{huang2004,egami2004,egami2007}, and $R$-band optical imaging with the 
8-m Subaru telescope to a $3\sigma$ depth of 27.9 mag \citep[in a 2$\arcsec$ aperture; 
see][]{ivison2004}. Finally, additional 
multi-frequency optical ($B,R,I,z$) imaging of the Lockman Hole has now been obtained
with the Subaru 8-m telescope, and a $K$-band image is now available from 
Deep Extragalactic Survey component of the 
UKIRT Deep Infrared Sky Survey (see Dye et al. 2008 for details).

\label{sec:obs}

\begin{figure}
\epsfig{figure=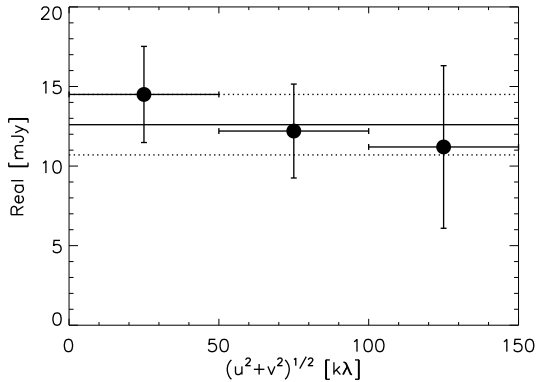,width=79mm}
\caption{The real visibility amplitudes versus projected baseline
($(u^2+v^2)^{1/2}$), centred on the position of LH\,850.02.  They
indicate that the submm emission is not resolved by the SMA out to
$\sim$120\,k$\lambda$.  This sets an upper limit on the angular scale
of the source to $\lsim$1\,arcsec.}
\label{fig:vis}
\end{figure} 

The target, LH\,850.02, is the brightest SMG in the SHADES Lockman
Hole sample (deboosted flux density of $S_{\rm 850\mu m} = 13.4\pm
2.1$\,mJy) and was detected at $6.8\sigma$ significance with SCUBA
\citep{coppin2006}.  It was first discovered as a $5.7\sigma$ MAMBO
source \citep[$S_{\rm 1200\mu m} = 5.7\pm 1.0$\,mJy;][]{greve2004},
and was included in the \citet{ivison2005} sample of `robust' SMGs.
It is also the brightest 1100-\micron\ SHADES-AzTEC source in the area
covered by both SCUBA and MAMBO (Austermann et al. 2008, in prep.),
but is not detected in the extremely deep ($\sim$600\,ks) {\em
XMM/Newton} imaging of the LH \citep{brunner2008}.

There are two candidate 1.4-GHz radio continuum counterparts within
the SCUBA beam identified by \citet{ivison2007}: a northern candidate
with flux density $S_{\rm 1.4GHz} = 40.7\pm 5.6$\,\microjy\
(LH\,850.02N: 3.5$\arcsec$ offset, $P=0.97$ probability of association), and a
southern candidate with $S_{\rm 1.4GHz} = 52.4\pm 5.2$\,\microjy\
(LH\,850.02S: 3.5$\arcsec$ offset, $P=0.98$). Imaging at 610\,MHz with the Giant
Metre-wave Radio Telescope (Ibar et al., in prep.) fails to
separate the two 1.4-GHz emitters, but shows a clear detection with a
peak flux density of 140\,\microjyend\,beam$^{-1}$ and total
integrated flux of 277\,\microjyend. In a matched-resolution 1.4-GHz
image, we find a source with a peak flux density of
57\,\microjyend\,beam$^{-1}$ and total integrated flux of
97\,\microjy, making LH\,850.02 a fairly-steep-spectrum radio source
($\alpha = -1.3$); marginally consistent with a starburst
\citep[$\alpha\approx -0.8$;][]{condon1992}, and possibly analogous to
high-redshift radio galaxies \citep[e.g.,][]{DeBreuk2000}, though
orders of magnitude fainter.

Finally, we can exploit multi-wavelength photometry (see Table~1) of
the LH\,850.02 system to gain some insight into the nature of SMGs
with multiple radio counterparts, which make up $\sim$10 per cent of
the SMG population \citep{ivison2002,pope2006} and $\sim$20 per cent
of radio-identified SMGs \citep{ivison2007}.  Photometric
redshifts derived from optical and near-IR photometry \citep[see
Figure~\ref{fig:photoz};][]{dye2008} indicate that LH\,850.02N
($z=3.41^{+0.10}_{-0.15}$) and LH\,850.02S ($z=3.23^{+0.11}_{-0.47}$)
may be physical associated. Furthermore, it is unlikely that LH850.02N and S are
duplicate images of a strongly lensed object because they have 
very different mid--IR SEDs -- LH850.02S has 24--\micron\ emission -- 
and their substantial relative separation ($\sim 6\arcsec$) would require a very 
strong lensing potential \citep[see e.g.,][]{grogin1996,keeton2000}.

While the slightly more northerly position of the MAMBO detection
favours LH\,850.02N \citep{ivison2005}, statistical arguments and a
strong MIPS detection favour LH\,850.02S
\citep[e.g.,][]{ashby2006,pope2006}; these ambiguous multi-wavelength
counterparts make LH\,850.02 a compelling target for high-resolution
submm imaging.  As we will make use of it later in this work, we
present the multiwavelength photometry for both LH\,850.02N and
LH\,850.02S in Table~1. \\ \\

\begin{table*}
\begin{tabular}{ccc}
\hline
\hline
Band & LH\,850.02N & LH\,850.02S \\
& [$\mu$Jy] & [$\mu$Jy] \\
\hline
B & $<0.04$ & $0.72\pm0.01$\\ 
R & $0.11\pm0.01$ & $0.29\pm0.02$ \\
I & $0.16\pm0.01$ & $0.44\pm0.03$ \\
z & $0.07\pm0.03$ & $0.55\pm0.04$ \\
K & $<7.0$ & $<7.0$ \\
3.6$\mu$m & $6.3\pm0.18$ & $6.6\pm0.17$ \\
4.5$\mu$m & $10.2\pm2.3$ & $9.8\pm2.3$ \\
5.8$\mu$m &  $9.1\pm4.6$ & $10.2\pm4.9$ \\
8.0$\mu$m & $11.5\pm4.8$ & $11.1\pm4.6$ \\
24$\mu$m & $<55$ & $545\pm31$ \\
850$\mu$m & $13400\pm2400$ & \ldots \\
890$\mu$m & $12800\pm2000$ & $<6000$ \\
1200$\mu$m & $5700\pm1000$ & \ldots \\
20cm & $40.7\pm5.6$ & $52.4\pm5.2$ \\
\hline
\hline
\end{tabular}
\caption{Optical-radio photometry for the two faint galaxies associated with the alternative
radio counterparts to LH 850.2. Upper limits are quoted at the 3-$\sigma$ level.}
\end{table*}

\begin{figure*}
\begin{center}
\epsfig{figure=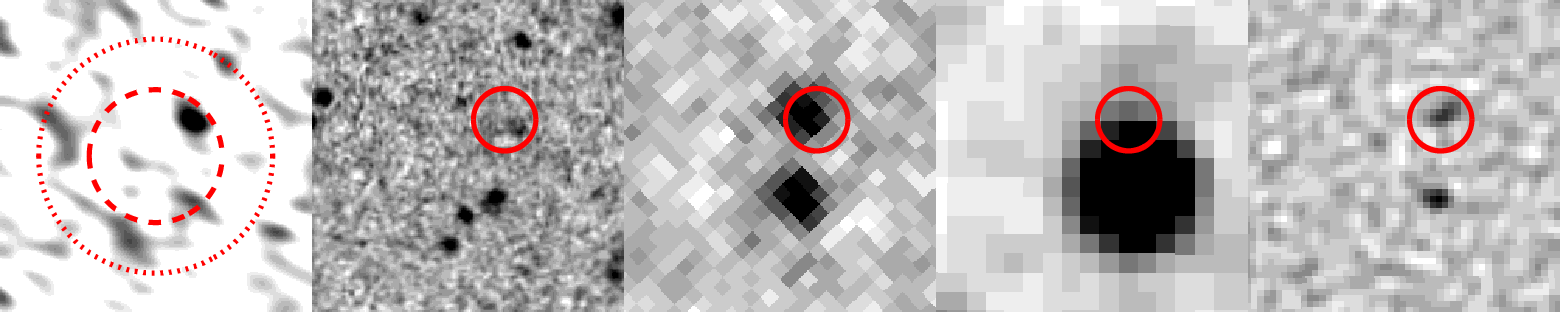,width=180mm}
\end{center}
\caption{Postage stamp images, centred on the SMA phase center position of
LH\,850.02, which is the original SCUBA centroid from \citet{coppin2006}.  
From left to right: SMA 890$\mu$m, Subaru $R$-band, IRAC 3.6-\micron,
MIPS 24-\micron\ and VLA 1.4-GHz imaging.  The
red dotted line indicates the {\sc fwhm} of the SCUBA beam, and the
red dashed line indicates the $2\sigma$ positional uncertainty
\citep{ivison2007}.  The synthesized beam size is 2.3\,arcsec $\times$ 1.3\,arcsec with a position angle of 60$^\circ$.  The red circle is 4-arcsec in diameter, roughly twice the SMA beam size, and the
stamps are 20\,arcsec on a side.  The SMA clearly identifies one compact point source as
the origin of the submm emission.}
\label{fig:stamps}
\end{figure*} 

\section{Observations and Data Reduction}
\label{sec:obs}

The SMA observations were performed during 2007 March in the compact
configuration (beam size $\sim$2\,arcsec) in excellent weather ($\tau_{\rm 225GHz}\lsim 0.08$) 
with a total on-source integration time of approximately 6\,hr. The USB was tuned to 345 GHz, and 
combined with the LSB for an effective bandwidth of $\sim 4$ GHz at 340 GHz, which yielded a final synthesised image r.m.s.\ of 1.95\,mJy.  The pointing centre was the
original SCUBA position from \citet{coppin2006}:
$\rm\alpha(\textrm{J2000}) = 10^h52^m57.32^s$ and
$\rm\delta(\textrm{J2000}) = +57^\circ21'05.8\arcsec$.  The data were
calibrated using the {\sc mir} software package \citep{scoville1993},
modified for the SMA.  Passband calibration was done using 3C\,84,
3C\,111, and Callisto.  The absolute flux scale was set using observations of Callisto and is estimated to be accurate to better than 20\%.  Time-dependent complex gain calibration was
done using 0958+655 (0.6\,Jy, 21.8$^{\circ}$ away) and 0927+390
(1.8\,Jy, 37.7$^{\circ}$ away).  The calibrator 0958+655 was also calibrated
independently using 0927+390 and used
for empirical verification of the astrometric uncertainty and the
angular size of the target. Positions and flux densities were derived
from the calibrated visibilities using the {\sc miriad}
\citep{sault1995} software package.

We detect LH\,850.02 in the synthesized image at $\gsim 6\sigma$.  The
calibrated visibilities were best fit by a single point source (see
Figures~\ref{fig:vis} and \ref{fig:stamps}) with an integrated flux
density of $S_{\rm 890\mu m} = 12.8\pm 2.0$\,mJy at a position of
$\rm\alpha(\textrm{J2000}) = 10^h52^m57.162^s$ and
$\rm\delta(\textrm{J2000}) = +57^\circ21'07.97\arcsec$, which is offset
from the original SCUBA position by 3.28\,arcsec.  The astrometric
uncertainties are $\Delta\alpha=0.24$\,arcsec (0.20\,arcsec
systematic; 0.13\,arcsec statistical) and $\Delta\delta=0.22$\,arcsec
(0.19\,arcsec systematic; 0.10\,arcsec statistical), in agreement with
the expectations of \citet{downes1999} and \citet{younger2007}.

\section{Discussion}
\label{sec:discuss}

The synthesised SMA image (see Figure~\ref{fig:stamps}) clearly shows a
single point source within the SCUBA beam. As noted in
\S~\ref{sec:obs}, this source is offset from the SCUBA centroid by
3.28\,arcsec, as compared to the $1\sigma$ SCUBA positional uncertainty of
2.1\,arcsec \citep[$\sigma \sim 0.91\,\theta/(\textrm{S/N})$, where
$\theta$ is the SCUBA beam {\sc fwhm} of 14\,arcsec and S/N is the
signal-to-noise ratio, corrected for flux boosting;][]{ivison2007}.
The point-source fit to the visibility data ($S_{\rm 890\mu m} =
12.8\pm2.0$\,mJy) is perfectly consistent with the expectation from
the deboosted SCUBA flux of $S_{\rm 850\mu m} = 13.4\pm 2.1$\,mJy
 ($13.1\pm2.3$ mJy when centered on the SMA position),
assuming a reasonable range of spectral slopes and temperatures 
($S_\nu\propto B_\nu(T) \nu^\beta$; $\beta \approx 1-2$, $T\approx 20-60$ {\sc k}).
This supports the compactness of the submm emission seen in the 
visibility function (see Figure~\ref{fig:vis}); virtually none of the flux seen in the 
SCUBA map has been resolved out.

In Figure~\ref{fig:stamps}, we present (from left to right) the SMA ``dirty" map 
and the derived submm position overlaid on optical ($R$-band), IRAC 3.6-\micronend, MIPS
24-\micron\ and VLA 1.4-GHz imaging data.  The SMA image clearly
singles out the weaker of the two candidate radio counterparts, and this is the one which
is {\it not} associated with the bright 24-\micron\ source\footnote{There is a $\sim 2\sigma$ peak nearly coincident with LH850.02S; the
$3\sigma$ upper limit is listed in Table~1.  However we note that the primary detection accounts entirely for the SCUBA flux.}.  
This is consistent with 
the photometric redshift of LH850.02N if we assume a starburst--dominated 
mid-IR spectrum similar to local ultraluminous IR galaxies
\citep[ULIRGs; e.g.][]{armus2007,desai2007}, analogous $z\sim 2$
systems \citep[e.g.,][Huang et al. 2008, in prep.]{weedman2006,farrah2008}, or other SMGs \citep{lutz2005,valiante2007,menendez2007}.

The SCUBA and SMA imaging data -- including upper limits at the location of LH850.02S -- are marginally consistent with the observed far--IR SEDs of two luminous starbursts at $z\sim 4$ (LH850.02N) and $z\sim 2$ (LH850.02S) respectively.  However, SMA imaging in combination with optical/near--IR photometric redshifts for both sources favors a scenario in which the submm emission from LH\,850.02 arises almost entirely from LH\,850.02N and is not a blend of two
lower-luminosity sources. This is consistent with the predicted rarity of SMGs
arising from confusion \citep{ivison2007}. At these redshifts, their 24\micronend--to--submm
ratios \citep[see also][]{wang2007} suggest that LH\,850.02N has an Arp\,220-type
starburst-dominated far-IR SED, while LH\,850.02S has a Mrk\,231-type
far-IR SED with a significant contribution from a warmer dust component such as a warm starburst \citep[OFRGs;][]{chapman2004b} or active nucleus. This is qualitatively similar to SMM J094303+4700 \citep{tacconi2006}, in which only one of the two radio counterparts \citep[H6: ][]{ledlow2002} shows strong CO emission, which could be explained by AGN heating of the dust in the intrinsically CO poor radio source (H7). Therefore, LH\,850.02 and, by analogy many other SMGs with multiple radio identifications, may be physically associated
systems in which the SMG starburst phase is associated with a period of intense AGN activity
\citep{sanders1996,page2004,hopkins2006}.

Furthermore, although the relatively low S/N ($\sim 6\sigma$) limits
the robustness of size measurements, 
the visibility function (Figure~\ref{fig:vis}) for LH\,850.02 
is consistent with a compact point--source out to $\sim$120\,k$\lambda$, 
from which we infer a maximum angular extent of $\lsim$1\,arcsec; similar to
other SMGs detected by SMA at 890\,\micron\
\citep{iono2006,younger2007,wang2007} and others observed at mm
wavelengths \citep{greve2005,tacconi2006}.  At a redshift of $z\approx 3-3.5$, 
this corresponds to a physical scale for the rest-frame far-IR
continuum of $\lsim$8\,kpc, consistent with a merger-driven starburst
analogous to local ULIRGs \citep[][see also C. Wilson et al. 2008, in prep.; Iono et al. 2008, in prep.]{downes1998,sakamato1999,
sakamato2006,iono2007}, and may be in conflict with cool, extended
cirrus dust models \citep{efstathiou2003,kaviani2003} or a monolithic
collapse scenario. Our constraints on size are only barely consistent
with extended ($\sim$1\,arcsec) starbursts of the kind inferred from
high-resolution radio imaging -- though some sources are reported as compact even at $\sim 0.2$ arcsec resolution \citep{chapman2004,biggs2007}.

\begin{figure}
\epsfig{figure=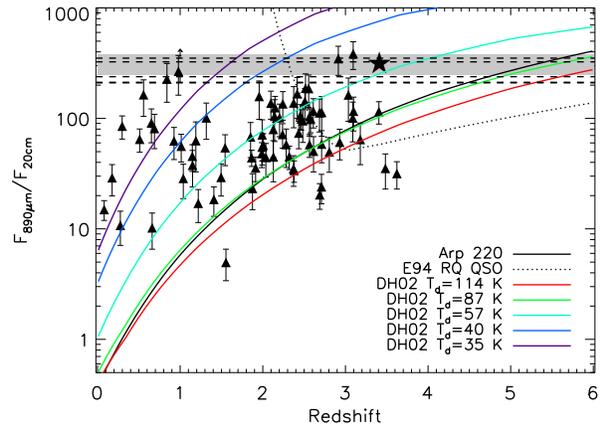,width=79mm}
\caption{The submm-to-radio flux density ratio versus redshift for
LH\,850.02 (grey region), and possible high-redshift SMGs (dashed
lines) from \citet{younger2007}.  For comparison, we show the model
track for Arp\,220 (black line), and the models of \citet{dale2002}
for a range of dust temperatures: $T_{\rm dust} =$ 114, 87, 57, 40,
and 35\,{\sc k}, where $T_{dust}$ is inferred from the $F_{60\mu m}/F_{100\mu m}$
flux density ratio.  The shaded region shows the observed flux density ratio of LH850.02N, and the star indicates its photometric redshift.  The black triangles are SMGs with optical redshifts
from \citet{chapman2005}.  Also shown is a model track (dotted line)
for the median radio-quiet quasar from \citet{elvis1994}.}
\label{fig:smmradio}
\end{figure} 

We can then use the observed submm--to--radio flux density ratio, coupled with 
photometric redshift and observed constraints on the physical scale of the rest--frame
far--IR to infer some of the physical properties of the starburst.  Figure~\ref{fig:smmradio}
shows the submm-to-radio flux density ratio $S_{\rm 890\mu m}/S_{\rm
1.4GHz}$ for LH\,850.02, as compared to the possible high-redshift
SMGs from \citet{younger2007}, SMGs with optical redshifts from
\citet{chapman2005}, tracks for Arp\,220, the models of
\citet{dale2002} and the median radio-quiet quasar spectral energy
distribution from \citet{elvis1994}.  The observed submm--to--radio flux density ratio of 
LH850.02N indicates a dust temperature of $\sim 60$ K, which is consistent with the 
observed compactness of the submm emission \citep[see][]{kaviani2003}.  Using the 
template SEDs of \citet{dale2002} for this best--fit dust temperature at the photometric 
redshift of $z\approx 3.3$, we find that the observed submm emission of $S_{890\mu m} = 12.8
\, {\rm mJy}$ 
corresponds to a total IR luminosity of $L \sim 2\times 10^{13}\, {\rm L_{\odot}}$, which -- 
assuming a \citet{salpeter1955} IMF -- indicates a 
total SFR of $\sim 3000\, M_\odot$ yr$^{-1}$ \citep{kennicutt1998cal,bell2003}.  
Should even higher-resolution imaging place tighter limits on the physical \
scale of the submm emission in LH850.02N, such a confined, luminous starburst 
may have important physical consequences; Eddington arguments \citep[e.g.,][]{murray2005,thompson2005} suggest a minimal scale for such regions.

Finally, considering its far--IR luminosity and compactness, the lack of detectable 
24\micron\ emission, and its 
observed submm--to--radio flux density ratio, we find that LH\,850.02 appears to share many of the
observed characteristics of other putative high-redshift
SMGs \citep{iono2006,younger2007,wang2007,dannerbauer2008}.  As discussed in more detail in
\citet{younger2007}, the existence of such a population provides tight
constraints on models of galaxy formation and evolution, and on dust
production. In addition, it fuels speculation that the brightest SMGs
may be the most distant \citep[see also][]{ivison2002} and suggests
significant and rapid down-sizing \citep[e.g.,][]{cowie1996} in the
SMG population over a relatively short interval in cosmic time \citep[see also][]{wall2008}.

\section{Conclusion}

We present high-resolution 890-\micron\ interferometric imaging of
LH\,850.02, a bright SMG in the LH, with the SMA. We detect
LH\,850.02 at $\gsim 6\sigma$ as a single compact (size
$\lsim$1\,arcsec, or $\lsim$8\,kpc) point source, and determine a
position accurate to $\sim$0.2\,arcsec.  From this, we identify
multi-wavelength counterparts and find that only one (LH850.02N) of two candidate
radio counterparts is associated with the submm emission. The nearby radio source with
strong 24-\micron\ emission (LH850.02S) does {\em not} contribute significantly to the submm flux.
In this case, and by analogy other SMGs with multiple radio counterparts, 
the radio continuum more reliably locates the source of the submm -- and with it
the properties of the associated starburst -- than does the mid--IR; similar to the results of \citet{younger2007}. Since both radio sources have similar photometric redshifts, and 
therefore may be physically associated, 
their respective 24\micronend--to--submm suggest that LH\,850.02N has an Arp\,220-type
starburst-dominated far-IR SED, while LH\,850.02S has a Mrk\,231-type
SED with a significant mid-infrared contribution from an active
nucleus.  

As a result of the relatively shallow, wide-field surveys conducted to
date with the AzTEC camera -- wherein objects of this type were first
found in significant numbers \citep{younger2007} -- existing
1100-\micron\ samples are typically $\gsim$2$\times$ brighter than
samples selected with SCUBA or MAMBO. We suggest that the recent
prevalence of candidate high-redshift SMGs is more
closely related to their high flux densities than to the long survey
wavelength, a tendency noted by \citet{ivison2002} and \citet{wall2008}.
It is therefore perhaps not surprising that we have here 
constrained the brightest source in the SHADES survey of the LH,
LH\,850.02, to lie at high redshift $z\gsim 3$.

\section*{Acknowledgements}
The SMA is a joint project between the Smithsonian Astrophysical
Observatory and the Academia Sinica Institute of Astronomy and
Astrophysics and is funded by the Smithsonian Institution and the
Academia Sinica.  This work is based in on observations made with the
{\it Spitzer Space Telescope}, which is operated by the Jet Propulsion
Laboratory, California Institute of Technology, under contract 1407,
and data collected at Subaru Telescope, which is operated by the
National Astronomical Observatory of Japan.  The James Clerk Maxwell
Telescope is operated by The Joint Astronomy Centre on behalf of the
Science and Technology Facilities Council of the United Kingdom, the
Netherlands Organisation for Scientific Research, and the National
Research Council of Canada.  IS acknowledges support from the 
Royal Society.

\bibliographystyle{mn2e}
\bibliography{../../smg}

\begin{thebibliography}{}

\bibitem[\protect\citeauthoryear{{Armus} et~al.,}{{Armus}
  et~al.}{2007}]{armus2007}
{Armus} L.,  et~al., 2007, \apj, 656, 148

\bibitem[\protect\citeauthoryear{{Ashby} et~al.,}{{Ashby}
  et~al.}{2006}]{ashby2006}
{Ashby} M.~L.~N.,  et~al., 2006, \apj, 644, 778

\bibitem[\protect\citeauthoryear{{Barger}, {Cowie} \& {Sanders}}{{Barger}
  et~al.}{1999}]{barger1999}
{Barger} A.~J.,  {Cowie} L.~L.,    {Sanders} D.~B.,  1999, \apjl, 518, L5

\bibitem[\protect\citeauthoryear{{Barger} et~al.,}{{Barger}
  et~al.}{1998}]{barger1998}
{Barger} A.~J.,  et~al., 1998, \nat, 394, 248

\bibitem[\protect\citeauthoryear{{Bell}}{{Bell}}{2003}]{bell2003}
{Bell} E.~F.,  2003, \apj, 586, 794

\bibitem[\protect\citeauthoryear{{Bertoldi} et~al.,}{{Bertoldi}
  et~al.}{2007}]{bertoldi2007}
{Bertoldi} F.,  et~al., 2007, \apjs, 172, 132

\bibitem[\protect\citeauthoryear{{Biggs} \& {Ivison}}{{Biggs} \&
  {Ivison}}{2006}]{biggs2006}
{Biggs} A.~D.,  {Ivison} R.~J.,  2006, MNRAS, 371, 963

\bibitem[\protect\citeauthoryear{{Biggs} \& {Ivison}}{{Biggs} \&
  {Ivison}}{2007}]{biggs2007}
{Biggs} A.~D.,  {Ivison} R.~J.,  2007, MNRAS, in press [astro-ph/0712.3047]

\bibitem[\protect\citeauthoryear{{Blain}, {Chapman}, {Smail} \&
  {Ivison}}{{Blain} et~al.}{2004}]{blain2004}
{Blain} A.~W.,  {Chapman} S.~C.,  {Smail} I.,    {Ivison} R.,  2004, \apj, 611,
  725

\bibitem[\protect\citeauthoryear{{Blain}, {Smail}, {Ivison} \& {Kneib}}{{Blain}
  et~al.}{1999}]{blain1999}
{Blain} A.~W.,  {Smail} I.,  {Ivison} R.~J.,    {Kneib} J.-P.,  1999, \mnras,
  302, 632

\bibitem[\protect\citeauthoryear{{Blain}, {Smail}, {Ivison}, {Kneib} \&
  {Frayer}}{{Blain} et~al.}{2002}]{blain2002}
{Blain} A.~W.,  {Smail} I.,  {Ivison} R.~J.,  {Kneib} J.-P.,    {Frayer} D.~T.,
   2002, \physrep, 369, 111

\bibitem[\protect\citeauthoryear{{Borys}, {Chapman}, {Halpern} \&
  {Scott}}{{Borys} et~al.}{2003}]{borys2003}
{Borys} C.,  {Chapman} S.,  {Halpern} M.,    {Scott} D.,  2003, \mnras, 344,
  385

\bibitem[\protect\citeauthoryear{{Borys}, {Chapman}, {Halpern} \&
  {Scott}}{{Borys} et~al.}{2002}]{borys2002}
{Borys} C.,  {Chapman} S.~C.,  {Halpern} M.,    {Scott} D.,  2002, \mnras, 330,
  L63

\bibitem[\protect\citeauthoryear{{Brunner} et~al.,}{{Brunner}
  et~al.}{2008}]{brunner2008}
{Brunner} H.,  et~al., 2008, \aap, 479, 283

\bibitem[\protect\citeauthoryear{{Chapman}, {Blain}, {Smail} \&
  {Ivison}}{{Chapman} et~al.}{2005}]{chapman2005}
{Chapman} S.~C.,  {Blain} A.~W.,  {Smail} I.,    {Ivison} R.~J.,  2005, \apj,
  622, 772

\bibitem[\protect\citeauthoryear{{Chapman}, {Scott}, {Borys} \&
  {Fahlman}}{{Chapman} et~al.}{2002}]{chapman2002}
{Chapman} S.~C.,  {Scott} D.,  {Borys} C.,    {Fahlman} G.~G.,  2002, \mnras,
  330, 92

\bibitem[\protect\citeauthoryear{{Chapman}, {Smail}, {Blain} \&
  {Ivison}}{{Chapman} et~al.}{2004}]{chapman2004b}
{Chapman} S.~C.,  {Smail} I.,  {Blain} A.~W.,    {Ivison} R.~J.,  2004, \apj,
  614, 671

\bibitem[\protect\citeauthoryear{{Chapman}, {Smail}, {Windhorst}, {Muxlow} \&
  {Ivison}}{{Chapman} et~al.}{2004}]{chapman2004}
{Chapman} S.~C.,  {Smail} I.,  {Windhorst} R.,  {Muxlow} T.,    {Ivison} R.~J.,
   2004, \apj, 611, 732

\bibitem[\protect\citeauthoryear{{Condon}}{{Condon}}{1992}]{condon1992}
{Condon} J.~J.,  1992, \araa, 30, 575

\bibitem[\protect\citeauthoryear{{Coppin} et~al.,}{{Coppin}
  et~al.}{2006}]{coppin2006}
{Coppin} K.,  et~al., 2006, \mnras, 372, 1621

\bibitem[\protect\citeauthoryear{{Cowie}, {Barger} \& {Kneib}}{{Cowie}
  et~al.}{2002}]{cowie2002}
{Cowie} L.~L.,  {Barger} A.~J.,    {Kneib} J.-P.,  2002, \aj, 123, 2197

\bibitem[\protect\citeauthoryear{{Cowie}, {Songaila}, {Hu} \& {Cohen}}{{Cowie}
  et~al.}{1996}]{cowie1996}
{Cowie} L.~L.,  {Songaila} A.,  {Hu} E.~M.,    {Cohen} J.~G.,  1996, \aj, 112,
  839

\bibitem[\protect\citeauthoryear{{Dale} \& {Helou}}{{Dale} \&
  {Helou}}{2002}]{dale2002}
{Dale} D.~A.,  {Helou} G.,  2002, \apj, 576, 159

\bibitem[\protect\citeauthoryear{{Dannerbauer} et~al.,}{{Dannerbauer}
  et~al.}{2002}]{dannerbauer2002}
{Dannerbauer} H.,  et~al., 2002, \apj, 573, 473

\bibitem[\protect\citeauthoryear{{Dannerbauer} et~al.,}{{Dannerbauer}
  et~al.}{2004}]{dannerbauer2004}
{Dannerbauer} H.,  et~al., 2004, \apj, 606, 664

\bibitem[\protect\citeauthoryear{{Dannerbauer}, {Walter} \&
  {Morrison}}{{Dannerbauer} et~al.}{2008}]{dannerbauer2008}
{Dannerbauer} H.,  {Walter} F.,    {Morrison} G.,  2008, \apjl, 673, L127

\bibitem[\protect\citeauthoryear{{De Breuck}, {van Breugel}, {R{\"o}ttgering}
  \& {Miley}}{{De Breuck} et~al.}{2000}]{DeBreuk2000}
{De Breuck} C.,  {van Breugel} W.,  {R{\"o}ttgering} H.~J.~A.,    {Miley} G.,
  2000, \aaps, 143, 303

\bibitem[\protect\citeauthoryear{{Desai} et~al.,}{{Desai}
  et~al.}{2007}]{desai2007}
{Desai} V.,  et~al., 2007, \apj, 669, 810

\bibitem[\protect\citeauthoryear{{Downes} et~al.,}{{Downes}
  et~al.}{1999}]{downes1999}
{Downes} D.,  et~al., 1999, \aap, 347, 809

\bibitem[\protect\citeauthoryear{{Downes} \& {Solomon}}{{Downes} \&
  {Solomon}}{1998}]{downes1998}
{Downes} D.,  {Solomon} P.~M.,  1998, \apj, 507, 615

\bibitem[\protect\citeauthoryear{{Dunlop}}{{Dunlop}}{2001}]{dunlop2001}
{Dunlop} J.~S.,  2001, New Astronomy Reviews, 45, 609

\bibitem[\protect\citeauthoryear{{Dunlop}}{{Dunlop}}{2005}]{dunlop2005}
{Dunlop} J.~S.,  2005, in {de Grijs} R.,  {Gonz{\'a}lez Delgado} R.~M.,  eds,
  Starbursts: From 30 Doradus to Lyman Break Galaxies Vol.~329 of Astrophysics
  and Space Science Library, {SHADES: The Scuba HAlf Degree Extragalactic
  Survey}.
pp 121--+

\bibitem[\protect\citeauthoryear{{Dwek} et~al.,}{{Dwek}
  et~al.}{1998}]{dwek1998}
{Dwek} E.,  et~al., 1998, \apj, 508, 106

\bibitem[\protect\citeauthoryear{{Dye} et~al.,}{{Dye}  et~al.}{2008}]{dye2008}
{Dye} S.,  et~al., 2008, MNRAS, [astro-ph/0802.0497]

\bibitem[\protect\citeauthoryear{{Eales} et~al.,}{{Eales}
  et~al.}{1999}]{eales1999}
{Eales} S.,  et~al., 1999, \apj, 515, 518

\bibitem[\protect\citeauthoryear{{Eales} et~al.,}{{Eales}
  et~al.}{2000}]{eales2000}
{Eales} S.,  et~al., 2000, \aj, 120, 2244

\bibitem[\protect\citeauthoryear{{Efstathiou} \& {Rowan-Robinson}}{{Efstathiou}
  \& {Rowan-Robinson}}{2003}]{efstathiou2003}
{Efstathiou} A.,  {Rowan-Robinson} M.,  2003, \mnras, 343, 322

\bibitem[\protect\citeauthoryear{{Egami} et~al.,}{{Egami}
  et~al.}{2004}]{egami2004}
{Egami} E.,  et~al., 2004, \apjs, 154, 130

\bibitem[\protect\citeauthoryear{{Egami} et~al.,}{{Egami}
  et~al.}{2007}]{egami2007}
{Egami} E.,  et~al., 2007, in preparation

\bibitem[\protect\citeauthoryear{{Elvis} et~al.,}{{Elvis}
  et~al.}{1994}]{elvis1994}
{Elvis} M.,  et~al., 1994, \apjs, 95, 1

\bibitem[\protect\citeauthoryear{{Farrah} et~al.,}{{Farrah}
  et~al.}{2008}]{farrah2008}
{Farrah} D.,  et~al., 2008, ApJ, in press [astro-ph/0801.1842]

\bibitem[\protect\citeauthoryear{{Fazio} et~al.,}{{Fazio}
  et~al.}{2004}]{fazio2004}
{Fazio} G.~G.,  et~al., 2004, \apjs, 154, 10

\bibitem[\protect\citeauthoryear{{Fixsen} et~al.,}{{Fixsen}
  et~al.}{1998}]{fixsen1998}
{Fixsen} D.~J.,  et~al., 1998, \apj, 508, 123

\bibitem[\protect\citeauthoryear{{Frayer}, {Smail}, {Ivison} \&
  {Scoville}}{{Frayer} et~al.}{2000}]{frayer2000}
{Frayer} D.~T.,  {Smail} I.,  {Ivison} R.~J.,    {Scoville} N.~Z.,  2000, \aj,
  120, 1668

\bibitem[\protect\citeauthoryear{{Genzel} et~al.,}{{Genzel}
  et~al.}{2003}]{genzel2003}
{Genzel} R.,  et~al., 2003, \apj, 584, 633

\bibitem[\protect\citeauthoryear{{Greve} et~al.,}{{Greve}
  et~al.}{2004}]{greve2004}
{Greve} T.~R.,  et~al., 2004, \mnras, 354, 779

\bibitem[\protect\citeauthoryear{{Greve} et~al.,}{{Greve}
  et~al.}{2005}]{greve2005}
{Greve} T.~R.,  et~al., 2005, \mnras, 359, 1165

\bibitem[\protect\citeauthoryear{{Grogin} \& {Narayan}}{{Grogin} \&
  {Narayan}}{1996}]{grogin1996}
{Grogin} N.~A.,  {Narayan} R.,  1996, \apj, 464, 92

\bibitem[\protect\citeauthoryear{{Hauser} et~al.,}{{Hauser}
  et~al.}{1998}]{hauser1998}
{Hauser} M.~G.,  et~al., 1998, \apj, 508, 25

\bibitem[\protect\citeauthoryear{{Ho}, {Moran} \& {Lo}}{{Ho}
  et~al.}{2004}]{ho2004}
{Ho} P.~T.~P.,  {Moran} J.~M.,    {Lo} K.~Y.,  2004, \apjl, 616, L1

\bibitem[\protect\citeauthoryear{{Holland} et~al.,}{{Holland}
  et~al.}{1999}]{holland1999}
{Holland} W.~S.,  et~al., 1999, \mnras, 303, 659

\bibitem[\protect\citeauthoryear{{Hopkins} et~al.,}{{Hopkins}
  et~al.}{2006}]{hopkins2006}
{Hopkins} P.~F.,  et~al., 2006, \apjs, 163, 1

\bibitem[\protect\citeauthoryear{{Huang} et~al.,}{{Huang}
  et~al.}{2004}]{huang2004}
{Huang} J.-S.,  et~al., 2004, \apjs, 154, 44

\bibitem[\protect\citeauthoryear{{Hughes} et~al.,}{{Hughes}
  et~al.}{1998}]{hughes1998}
{Hughes} D.~H.,  et~al., 1998, \nat, 394, 241

\bibitem[\protect\citeauthoryear{{Iono} et~al.,}{{Iono}
  et~al.}{2006}]{iono2006}
{Iono} D.,  et~al., 2006, \apjl, 640, L1

\bibitem[\protect\citeauthoryear{{Iono} et~al.,}{{Iono}
  et~al.}{2007}]{iono2007}
{Iono} D.,  et~al., 2007, \apj, 659, 283

\bibitem[\protect\citeauthoryear{{Ivison} et~al.,}{{Ivison}
  et~al.}{2002}]{ivison2002}
{Ivison} R.~J.,  et~al., 2002, \mnras, 337, 1

\bibitem[\protect\citeauthoryear{{Ivison} et~al.,}{{Ivison}
  et~al.}{2004}]{ivison2004}
{Ivison} R.~J.,  et~al., 2004, \apjs, 154, 124

\bibitem[\protect\citeauthoryear{{Ivison} et~al.,}{{Ivison}
  et~al.}{2005}]{ivison2005}
{Ivison} R.~J.,  et~al., 2005, \mnras, 364, 1025

\bibitem[\protect\citeauthoryear{{Ivison} et~al.,}{{Ivison}
  et~al.}{2007}]{ivison2007}
{Ivison} R.~J.,  et~al., 2007, \mnras, 380, 199

\bibitem[\protect\citeauthoryear{{Kaviani}, {Haehnelt} \&
  {Kauffmann}}{{Kaviani} et~al.}{2003}]{kaviani2003}
{Kaviani} A.,  {Haehnelt} M.~G.,    {Kauffmann} G.,  2003, \mnras, 340, 739

\bibitem[\protect\citeauthoryear{{Keeton} et~al.,}{{Keeton}
  et~al.}{2000}]{keeton2000}
{Keeton} C.~R.,  et~al., 2000, \apj, 542, 74

\bibitem[\protect\citeauthoryear{{Kennicutt}
  Jr.}{{Kennicutt}}{1998}]{kennicutt1998cal}
{Kennicutt} Jr. R.~C.,  1998, \araa, 36, 189

\bibitem[\protect\citeauthoryear{{Kneib} et~al.,}{{Kneib}
  et~al.}{2005}]{kneib2005}
{Kneib} J.-P.,  et~al., 2005, \aap, 434, 819

\bibitem[\protect\citeauthoryear{{Knudsen} et~al.,}{{Knudsen}
  et~al.}{2006}]{knudsen2006}
{Knudsen} K.~K.,  et~al., 2006, \mnras, 368, 487

\bibitem[\protect\citeauthoryear{{Knudsen}, {van der Werf} \&
  {Kneib}}{{Knudsen} et~al.}{2008}]{knudsen2008}
{Knudsen} K.~K.,  {van der Werf} P.~P.,    {Kneib} J.~.,  2008, MNRAS, in press
  [astro-ph/0712.1904]

\bibitem[\protect\citeauthoryear{{Laurent} et~al.,}{{Laurent}
  et~al.}{2005}]{laurent2005}
{Laurent} G.~T.,  et~al., 2005, \apj, 623, 742

\bibitem[\protect\citeauthoryear{{Ledlow} et~al.,}{{Ledlow}
  et~al.}{2002}]{ledlow2002}
{Ledlow} M.~J.,  et~al., 2002, \apjl, 577, L79

\bibitem[\protect\citeauthoryear{{Lutz} et~al.,}{{Lutz}
  et~al.}{2005}]{lutz2005}
{Lutz} D.,  et~al., 2005, \apjl, 625, L83

\bibitem[\protect\citeauthoryear{{Men{\'e}ndez-Delmestre}
  et~al.,}{{Men{\'e}ndez-Delmestre}  et~al.}{2007}]{menendez2007}
{Men{\'e}ndez-Delmestre} K.,  et~al., 2007, \apjl, 655, L65

\bibitem[\protect\citeauthoryear{{Mortier} et~al.,}{{Mortier}
  et~al.}{2005}]{mortier2005}
{Mortier} A.~M.~J.,  et~al., 2005, \mnras, 363, 563

\bibitem[\protect\citeauthoryear{{Murray}, {Quataert} \& {Thompson}}{{Murray}
  et~al.}{2005}]{murray2005}
{Murray} N.,  {Quataert} E.,    {Thompson} T.~A.,  2005, \apj, 618, 569

\bibitem[\protect\citeauthoryear{{Neri} et~al.,}{{Neri}
  et~al.}{2003}]{neri2003}
{Neri} R.,  et~al., 2003, \apjl, 597, L113

\bibitem[\protect\citeauthoryear{{Page}, {Stevens}, {Ivison} \&
  {Carrera}}{{Page} et~al.}{2004}]{page2004}
{Page} M.~J.,  {Stevens} J.~A.,  {Ivison} R.~J.,    {Carrera} F.~J.,  2004,
  \apjl, 611, L85

\bibitem[\protect\citeauthoryear{{Pei}, {Fall} \& {Hauser}}{{Pei}
  et~al.}{1999}]{pei1999}
{Pei} Y.~C.,  {Fall} S.~M.,    {Hauser} M.~G.,  1999, \apj, 522, 604

\bibitem[\protect\citeauthoryear{{Pope} et~al.,}{{Pope}
  et~al.}{2006}]{pope2006}
{Pope} A.,  et~al., 2006, \mnras, 370, 1185

\bibitem[\protect\citeauthoryear{{Rieke},  et~al.,}{{Rieke}
  et~al.}{2004}]{rieke2004}
{Rieke} G.~H.,     et~al., 2004, \apjs, 154, 25

\bibitem[\protect\citeauthoryear{{Sakamoto} et~al.,}{{Sakamoto}
  et~al.}{1999}]{sakamato1999}
{Sakamoto} K.,  et~al., 1999, \apj, 514, 68

\bibitem[\protect\citeauthoryear{{Sakamoto}, {Ho} \& {Peck}}{{Sakamoto}
  et~al.}{2006}]{sakamato2006}
{Sakamoto} K.,  {Ho} P.~T.~P.,    {Peck} A.~B.,  2006, \apj, 644, 862

\bibitem[\protect\citeauthoryear{{Salpeter}}{{Salpeter}}{1955}]{salpeter1955}
{Salpeter} E.~E.,  1955, \apj, 121, 161

\bibitem[\protect\citeauthoryear{{Sanders} \& {Mirabel}}{{Sanders} \&
  {Mirabel}}{1996}]{sanders1996}
{Sanders} D.~B.,  {Mirabel} I.~F.,  1996, \araa, 34, 749

\bibitem[\protect\citeauthoryear{{Sault}, {Teuben} \& {Wright}}{{Sault}
  et~al.}{1995}]{sault1995}
{Sault} R.~J.,  {Teuben} P.~J.,    {Wright} M.~C.~H.,  1995, in {Shaw} R.~A.,
  {Payne} H.~E.,   {Hayes} J.~J.~E.,  eds, ASP Conf. Ser. 77: Astronomical Data
  Analysis Software and Systems IV {A Retrospective View of MIRIAD}.
p.~433

\bibitem[\protect\citeauthoryear{{Scott} et~al.,}{{Scott}
  et~al.}{2006}]{dscott2006}
{Scott} D.,  et~al., 2006, in Bulletin of the American Astronomical Society
  Vol.~38 of Bulletin of the American Astronomical Society, {A Deep AzTEC Map
  of the GOODS-North Field}.
pp 1072--+

\bibitem[\protect\citeauthoryear{{Scott} et~al.,}{{Scott}
  et~al.}{2008}]{scott2008}
{Scott} K.~S.,  et~al., 2008, MNRAS, in press [astro-ph/0801.2779]

\bibitem[\protect\citeauthoryear{{Scott} et~al.,}{{Scott}
  et~al.}{2002}]{scott2002}
{Scott} S.~E.,  et~al., 2002, \mnras, 331, 817

\bibitem[\protect\citeauthoryear{{Scoville}, {Carlstrom}, {Chandler},
  {Phillips}, {Scott}, {Tilanus} \& {Wang}}{{Scoville}
  et~al.}{1993}]{scoville1993}
{Scoville} N.~Z.,  {Carlstrom} J.~E.,  {Chandler} C.~J.,  {Phillips} J.~A.,
  {Scott} S.~L.,  {Tilanus} R.~P.~J.,    {Wang} Z.,  1993, \pasp, 105, 1482

\bibitem[\protect\citeauthoryear{{Serjeant} et~al.,}{{Serjeant}
  et~al.}{2003}]{serjeant2003}
{Serjeant} S.,  et~al., 2003, \mnras, 344, 887

\bibitem[\protect\citeauthoryear{{Sheth} et~al.,}{{Sheth}
  et~al.}{2004}]{sheth2004}
{Sheth} K.,  et~al., 2004, \apjl, 614, L5

\bibitem[\protect\citeauthoryear{{Smail}, {Ivison} \& {Blain}}{{Smail}
  et~al.}{1997}]{smail1997}
{Smail} I.,  {Ivison} R.~J.,    {Blain} A.~W.,  1997, \apjl, 490, L5

\bibitem[\protect\citeauthoryear{{Tacconi} et~al.,}{{Tacconi}
  et~al.}{2006}]{tacconi2006}
{Tacconi} L.~J.,  et~al., 2006, \apj, 640, 228

\bibitem[\protect\citeauthoryear{{Thompson}, {Quataert} \& {Murray}}{{Thompson}
  et~al.}{2005}]{thompson2005}
{Thompson} T.~A.,  {Quataert} E.,    {Murray} N.,  2005, \apj, 630, 167

\bibitem[\protect\citeauthoryear{{Valiante} et~al.,}{{Valiante}
  et~al.}{2007}]{valiante2007}
{Valiante} E.,  et~al., 2007, \apj, 660, 1060

\bibitem[\protect\citeauthoryear{{Wall}, {Pope} \& {Scott}}{{Wall}
  et~al.}{2008}]{wall2008}
{Wall} J.~V.,  {Pope} A.,    {Scott} D.,  2008, \mnras, 383, 435

\bibitem[\protect\citeauthoryear{{Wang}, {Cowie} \& {Barger}}{{Wang}
  et~al.}{2004}]{wang2004}
{Wang} W.-H.,  {Cowie} L.~L.,    {Barger} A.~J.,  2004, \apj, 613, 655

\bibitem[\protect\citeauthoryear{{Wang} et~al.,}{{Wang}
  et~al.}{2007}]{wang2007}
{Wang} W.-H.,  et~al., 2007, \apjl, 670, L89

\bibitem[\protect\citeauthoryear{{Webb} et~al.,}{{Webb}
  et~al.}{2003}]{webb2003}
{Webb} T.~M.,  et~al., 2003, \apj, 587, 41

\bibitem[\protect\citeauthoryear{{Weedman} et~al.,}{{Weedman}
  et~al.}{2006}]{weedman2006}
{Weedman} D.,  et~al., 2006, \apj, 653, 101

\bibitem[\protect\citeauthoryear{{Wilson} et~al.,}{{Wilson}
  et~al.}{2008}]{wilson2008}
{Wilson} G.~W.,  et~al., 2008, MNRAS, in press [astro-ph/0801.2783], 801

\bibitem[\protect\citeauthoryear{{Younger} et~al.,}{{Younger}
  et~al.}{2007}]{younger2007}
{Younger} J.~D.,  et~al., 2007, \apj, 671, 1531

\end{thebibliography}

\end{document}